# Unveiling and engineering of third order optical nonlinearities in $NiCo_2O_4$ nanoflowers

Ankit Sharma[1], Pritam Khan[2*], Dipendranath Mandal[1], Mansi Pathak,[3] C. S. Rout,[3] and K. V. Adarsh[1*]

[1] Department of Physics, Indian Institute of Science Education and Research, Bhopal, 462066 India

[2] Department of Physics and Bernal Institute, University of Limerick, Limerick V94 T9PX, Ireland

[3] Centre for Nano and Materials Science, Jain University, Ramanagaram, Bangalore, 562112, India

In this paper, we demonstrate $NiCo_2O_4$ (NCO) as an efficient new nonlinear optical material with straightforward potential applications in optical limiting devices. We obtain nonlinear absorption coefficient ($\beta$) and nonlinear refractive index ($n_2$) in parallel by performing Z-scan technique, in both open and closed aperture configurations, respectively. To understand the mechanism responsible for third order optical nonlinearity in NCO, we excited the sample with contrasting laser pulse durations, 7 ns and 120 fs, at two different off-resonant wavelengths. For ns excitation, nonlinearity is mediated by excited state absorption (ESA) and free-carrier absorption, that gives rise to large $\beta_{ESA}$ and positive $n_2$. On the other hand, when excited with fs laser, two-photon absorption (TPA) takes place and bound carriers induce strong negative $n_2$. The values obtained for $\beta$ and $n_2$ in NCO are found to be higher among the family of well-known conventional transition metal oxides, therefore are promising for optics and other photonics applications.

* Author to whom correspondence should be addressed:
Electronic mail: adarsh@iiserb.ac.in & Pritam.Khan@ul.ie

# 1. Introduction

Recently, there has been huge upsurge to find new materials with high third order optical nonlinearity for integration with complementary metal-oxide-semiconductor (CMOS) technology for all-optical on-a-chip device applications [1]. For example, developing next-generation high-speed optoelectronic devices, e.g., optical modulators [2], optical switches [3], optical limiters [4,5] etc. with materials having high nonlinear absorption and a fast response time. Although a plethora of materials including carbon nanotubes [6], graphene-based material [4], glasses [7], 2D materials [8] etc. have been tested, there exists a trade-off between their optical limiting strength, damage thresholds and response time. In this context, transition-metal oxides (TMOs) have emerged as potential candidate for CMOS-compatible materials thanks to their large third-order optical nonlinearities with fast response time [9-11]. Apart from that, TMOs also belong to the category of transparent conducting oxides (TCOs) that demonstrate excellent infrared (IR) transparency and high conductivity [12,13] with potential applications in optical switching and sensing.

Among the family of TMOs, Cobalt oxide ($Co_3O_4$, hereafter CoO) stands apart owing to its high optical nonlinearity along with good thermal and mechanical stability. Recent studies have shown that CoO exhibits strong third and fifth order nonlinearity depending on the size and pulse-duration of the excitation laser [10,14]. For a good TCO material, there is always a trade-off between high conductivity and high transparency, since they are cross reciprocal to each other. For example, n-type TCOs (ITO, IZO, $SnO_2$) provide high conductivity with limited transmittivity in the IR [15,16] while a p-type TCOs (CoO, Ni/NiO) demonstrate excellent IR transmittivity with week conductivity [17,18]. Recent studies have shown that NCO, mixed oxide films containing Co and Ni in a spinel crystal structure, can provide remarkable transmittivity over a broadband from visible to IR regime with up to five orders of higher conductivity than either of CoO or NiO [18,19]. Likewise, NCO has been used extensively in multimodal applications, e.g., electrocatalysis [20], optical sensor [21], solar cells [22] etc. In spite of such enticing optical properties and plethora of applications, the optical nonlinearity in NCO is hitherto unexplored.

In light of the above developments, here we studied the third order optical nonlinearities in NCO as potential candidate for CMOS compatible optical limiting device. To the best of our knowledge, this paper unveils for the first time remarkable nonlinear absorption coefficient and nonlinear refractive index in NCO at off-resonance wavelengths for both nanosecond and femtosecond laser excitation. Experimental observations of ESA and TPA are explained well on the basis of band structure of NCO. We demonstrate that strong and fast nonlinearity in NCO provides straightforward application for designing optical limiting devices.

## 2. Experimental methods

### 2.1 Synthesis

We have synthesized NCO by a facile hydrothermal method described in detail in our recent work [23]. Typically, mixture of 30 mM $Ni(NO_3)_2.6H_2O$ and 60 mM of $Co(NO_3)_2.6H_2O$ were dissolved in 30 ml of distilled water. Afterwards 100 mM of urea dissolved in 10 ml of distilled water was gradually added to the mixture and stirred vigorously for 45 minutes till we obtain a homogeneous pink coloured solution. The mixture was then transferred to 50 ml of Teflon lined stainless steel autoclave to carry out the hydrothermal reaction for 12 hrs. at 200ºC. The obtained product first filtering and washed with distilled water and ethanol, then vacuum dried for 12 hrs at 60ºC and finally annealed for 3 hours at 350ºC to get the final product.

### 2.2 Structural characterization

Surface morphology of the synthesized sample was obtained from field emission scanning electron microscopy (FESEM, JEOL JSM-7100F, JEOL Ltd., Singapore). We employed X-ray photoelectron spectroscopy (XPS, Thermo K Alpha+ spectrometer, Al Kα X-rays with

1486.6eV energy) under ultrahigh vacuum condition for compositional analysis. Raman spectroscopy was performed by a micro-Raman spectrometer (Renishaw inVia Raman microscope) using a 532 nm laser operated at very nominal power of 1 mW/cm$^2$ to avoid any photo damage.

**2.3 Third order nonlinear optical measurements**

We use standard Z- scan technique to obtain third order optical nonlinearity in NCO. Z-scan is a precise method based on self-focusing using a single excitation beam which is used to extract β and $n_2$. In a Z-scan set up, we measure the change in the transmission of the excitation laser beam in the far-field as a function of sample position Z along the propagation direction through the beam focus. In the present work, we employed open aperture (OA) and closed aperture (CA) Z-scan measurements without and with an aperture in front of a photodiode, respectively. OA Z-scan profile determines β while CA Z-scan curve provides information on $n_2$. A schematic diagram of OA and CA Z-scan set up is shown in Figure 3.

**Nanosecond Z-scan measurements**: In this case, we excite the sample with 532 nm, 7 ns pulses from the second harmonic of a Nd-YAG laser. A low repetition rate of 10 Hz is used to avoid heating effect and photo damage. The ns laser beam is focused on the sample by a convex lens of focal length 20 cm. The Rayleigh length ($Z_0$) and beam waist in our experiment were ~3 mm and ~ 20 µm, respectively.

**Femtosecond Z-scan measurements**: In this experiment, 800 nm, 120 fs pulses from the Ti-sapphire laser were used to excite the sample. The repetition rate of the laser is set at 1 kHz. For ultrafast laser, we used a 30 cm plano-convex lens used to focus the beam on the sample. In this case, $Z_0$ and beam waist were 12 mm and 55 μm, respectively.

## 3. Results and Discussion

### 3.1 Structure and compositional analysis

Figure 1(a)-(b) depicts the surface morphology of NCO measured by FESEM. We observed a flower-like nanostructure of NCO (Figure 1(a)) which consists of nanoneedles spread isotopically in all direction thanks to spherical symmetry. The FESEM images at higher magnifications reveals the porosity of the nanoneedles as shown in Figure 1(b). For chemical composition analysis via valence bands states, we performed the XPS measurements of NCO that reveals the presence of Ni 2$p$, Co 2$p$, and O 1$s$ species as shown in the survey spectrum in Figure 1(c). The Gaussian fitting is employed to deconvolute the binding energy spectrum of individual components. Valence band spectrum of Ni 2$p$ in Figure 1(d) exhibits several peaks: binding energies of 854.48 and 871.56 eV are assigned to $Ni^{2+}$ state while the other two peaks at 855.93 and 872.80 eV are attributed to $Ni^{3+}$ state [24]. Apart from that, we observed two satellite peaks in Ni 2$p$ spectrum at side of binding energies of Ni 2$p_{3/2}$ and Ni 2$p_{1/2}$. Emission spectra of Co 2$p$ is depicted in Figure 1 (e). The binding energy peaks at 780.86 and 796.41 eV belongs to $Co^{2+}$ state, whereas the peaks at 779.16 and 794.38 eV are assigned to Co 2$p_{3/2}$ and Co 2$p_{1/2}$, spin–orbit doublet of $Co^{3+}$ state [24]. The high-resolution O 1$s$ spectrum is displayed in Figure 1(f) with three binding energy peaks at 529.79, 531.5 and 532.89 eV which are associated with metal-oxygen bonding, hydroxyl (-OH) group, and moisture adsorbed oxygen, respectively [24]. The position of the all the binding energy peaks obtained from XPS spectra are summarized in Table 1.

### 3.2. Vibrational and optical band analysis

The Raman spectra of NCO is shown in Figure 2(a). Two strong peaks are observed at 461 and 649 cm$^{-1}$, which are assigned to $E_g$ and $A_{1g}$ vibrational modes of NCO. Apart from that two week peaks are observed at 190 and 521 cm$^{-1}$, which are ascribed to $F_{2g}$ vibrational mode [23].

$A_{1g}$ peak represents the vibrations of octahedral oxygen to ions associated with $Co^{3+}$, while $E_g$ and $F_{2g}$ peaks are combined vibrations of tetrahedral and octahedral oxygen atoms in lattice [25,26]. Figure 2(b) shows the optical absorption spectrum of NCO which is analyzed to elucidate the bandgap. Since NCO is a direct band gap material [24], we use the following Tauc equation to calculate the bandgap [27]:

$$(\alpha h\nu)^2 = (h\nu - E) \qquad (1)$$

Where, α, h, ν, and E are the absorption coefficient, Plank's constant, frequency of incident photon, and bandgap respectively. After the best fitting of our experimental data, the bandgap energy of NCO is found to be 1.67 and 3.24 eV (inset of Figure 2(b), which are in good agreement with previous report [24]. The appearance of two bandgap has been explained by the co-existence of high and low spin of cobalt ion in NCO. Electronic band structure of NCO consists of high spin $Co^{2+}$ at tetrahedral site and low spin $Co^{3+}$ and $Ni^{3+}$ at octahedral sites. The band structure is defined as O 2$p$ orbital as a valence band (VB) and Co 3$d$ and Ni 3$d$ as conduction band (CB). Upon optical excitation, 3 electronic transitions (two inter-band and one intra-band) can take place in NCO, e.g. (i) O 2$p$ to higher Co 3$d$-$e_g$ (or Ni 3$d$-$e_g$) spin state, (ii) O 2$p$ to lower Co 3$d$-$t_{2g}$ (or Ni 3$d$-$t_{2g}$) spin state, and (iii) Co 3$d$-$t_{2g}$ to Co 3$d$-$e_g$ (or from Ni 3$d$-$t_{2g}$ to Ni 3$d$-$e_g$) [24]. Bandgap energies of 3.24 and 1.67 eV correspond to the two inter-band transitions (i) and (ii), respectively.

**3.3 Nonlinear optical response**

**3.3.1 Nanosecond pumping**

Figure 4(a) shows the OA Z-scan traces of NCO performed with different intensities of 7 ns, 532 nm laser pulses. At the lowest intensity of 0.20 GW/cm$^2$, response is dominated by saturable absorption (SA) with a week contribution from reverse saturable absorption (RSA).

As the input intensity is gradually increased from 0.20 to 0.64 GW/cm$^2$, we observed a strong intensity-dependent RSA, while SA remains nearly constant.

To understand the origin of RSA, we calculated the ground state absorption cross-section ($\sigma_{gs}$) and excited state absorption cross section ($\sigma_{ex}$) using following equations [28, 29]:

$$\sigma_{gs} = \frac{-\log T_0}{NL} \qquad (2)$$

$$\sigma_{ex} = \frac{-\log T_{max}}{NL} \qquad (3)$$

where, $T_0$, $T_{max}$, N, and L are linear transmission, saturated transmission at high intensity, ground state carrier density (sample concentration x Avogadro's number) and thickness of sample (or cuvette) respectively. Based on these equations, we found that $\sigma_{gs} = 4.63 \times 10^{-19}$ cm$^2$ and $\sigma_{ex} = 8.33 \times 10^{-19}$ cm$^2$. Since $\sigma_{ex}$ is higher than $\sigma_{gs}$, i.e., the ratio $\sigma_{ex}/\sigma_{gs}$ (= 1.8) is greater than unity, we conclude that RSA in NCO is originated from excited state absorption (ESA). The physical mechanism of ESA in NCO can be understood from the energy band diagram as shown in Figure 4(b). When excited with 532 nm (2.33 eV) laser, the photoexcited carriers first make an inter-band transition from VB to low spin state (Co 3$d$-t$_{2g}$ or Ni 3$d$-t$_{2g}$) in CB. With further increase in intensity, makes an intra-band transition from an already excited low spin state to high spin state (Co 3$d$-e$_g$ or Ni 3$d$-e$_g$) within the CB by ESA.

To quantify the ESA in NCO, we employed Z-scan theory based on propagation equation in the dispersion as a function of the position given by [30, 31],

$$\frac{dI}{dz} = -\alpha(I)I \qquad (4)$$

Where, I, z, and $\alpha(I)$ are the intensity, position, and intensity dependent absorption coefficient respectively.

$\alpha(I)$ is generally defined by the following equation:

$$\alpha(I) = \frac{\alpha_0}{1+\frac{I}{I_s}} + \beta_{ESA} I \qquad (5)$$

where $\alpha_0$, $I_s$, and are $\beta_{ESA}$ are the linear absorption coefficient, saturation intensity, and ESA coefficient, respectively. Eq. (4) and (5) provides us the intensity dependence $I_s$ and are $\beta_{ESA}$ which is plotted in Figure 4(c). We observed a significant enhancement of $\beta_{ESA}$ with input intensity in line with the theory of ESA. On the other hand, $I_s$ first decreases and then remains nearly constant with intensity. This observation is in line with the amplitude of SA, which is maximum at lowest intensity, then reduces and finally becomes intensity-independent. After the first observation of third order optical nonlinearity via ESA in NCO it quite enticing to compare $\beta_{ESA}$ with other TMO materials [32-36] and such comparison is shown in Table 2. Clearly, β in NCO is higher or comparable with other TMO materials that marks NCO as a strong candidate for nonlinear optics application. We envisage that the strong ESA in NCO has the potential to design optical limiting device. An optical limiter is a selective nonlinear device that allows to pass low-intensity laser beams while blocking high-intensity beams and plays an important role in protecting optoelectronic devices, e.g., photomultiplier tube, photodiode as well as human eye from intense laser [4, 5, 37]. To demonstrate direct application, we have plotted in Figure 4(d) the variation of output intensity ($I_{out}$) against the input intensity ($I_{in}$) at the peak intensity of 0.64 GW/cm$^2$. Clearly, at lower intensities ($I_{in} < 0.07$ GW/cm$^2$), $I_{out}$ scales a linear relationship with $I_{in}$ following Beer–Lambert law, thus the device remains inactive and allows the light beam to pass through. However, as $I_{in}$ increases, $I_{out}/I_{in}$ deviates from linearity as shown by the dashed line, i.e., the output intensity is blocked by NCO for all higher input intensities and the system acts an active optical limiting device.

After demonstrating strong ESA in NCO by OA Z-scan measurements, we performed CA Z-scan measurement to determine $n_2$. The normalized CA Z-scan trances of NCO in Figure 5 at 0.49 GW/cm$^2$ exhibits a valley-peak (V-P) structure, i.e., pre-focal transmission minimum (valley, $T_v$) is followed by a post-focal transmission maximum (peak, $T_p$). Such Z-scan traces results in positive $n_2$ associated with self-focussing effect. To calculate $n_2$, first we determine the on-axis phase shift which is related to difference of peak-valley from the following equation [31,37]:

$$\Delta\varphi = \frac{\Delta T_{\text{p-v}}}{0.406\,(1-S)^{0.25}} \tag{6}$$

Where, S is the ratio of transmitted intensity for close aperture to that of open aperture. This phase shift is related $\eta_2$ by following equation [31, 37],

$$n_2 = \frac{\Delta\varphi}{k\,I_0 L_{\text{eff}}} \tag{7}$$

Here, k is the wave vector ($2\pi/\lambda$), $I_0$ is the peak intensity at focus and $L_{\text{eff}}$ is the effective propagation length within the sample. From this calculation, we got nonlinear refractive index $\sim n_2 = 1.80 \times 10^{-4}$ cm$^2$/GW. Observation of positive $n_2$ can be attributed to free-carrier-induced effects since the strong intensity is high enough to generate free carriers.

### 3.3.2 Femtosecond pumping

OA Z-scan traces of NCO when excited with 120 fs, 800 nm laser is shown in Figure 6(a). At all excitation intensities transmission exhibits a dip at the focal point, therefore exhibits RSA. However, unlike ns excitation, transmission do not exhibit any significant change with intensity variation which indicates that RSA is arising from TPA which we will discuss soon in detail based on Figure 6(b). For 800 nm laser, excitation energy is 1.55 eV which is close to half the energy (3.24 eV) required for electronic transition for first inter-band transition between VB to high spin state in CB. In fact, this excitation energy is even less for low-energy inter-band transition energy between VB and low spin state in CB (1.67 eV). Therefore, the only plausible mechanism for the RSA observed in NCO with femtosecond laser is TPA. To quantify femtosecond laser induced TPA, we calculate the TPA coefficient ($\beta_{\text{TPA}}$) from Eqns. (3) and

(4) and found that it remains nearly constant ($3 \times 10^{-2}$ cm/GW) for all excitation intensities (Table 3), a signature of TPA. CA Z-scan traces for NCO under femtosecond excitation reveals a peak-valley shaped transmission curve, shown in Figure 6(c), which is typical for self-defocussing effect that will eventually results in negative $n_2$. Self-defocussing effect is known as negative lens effect in which Gaussian beam induces a lower refractive index at the center of the beam and maximum at the periphery that gives rise to negative $-n_2$. This observation indicates that sub-bandgap femtosecond excitation does not produce any free carriers, and since the thermal effects are negligible the observed nonlinearity is believed to be originated from bound carriers. Solving Eqns. (5) and (6) $n_2$ is found to be of the order of $10^{-7}$ cm$^2$/GW. Since $n_2$ decreases from large to low negative value with increase in intensity, it indicates that probability of generating free carriers is maximum at higher excitation intensities (Table 3). Remarkably, $\beta_{TPA}$ obtained for NCO for fs pulses is found to be highest as compared to other TMO materials as shown in Table 4 [38-39].

## 4. Conclusions

To the best of our knowledge, in this paper we unveiled for the first time third order optical nonlinearity in NCO by employing OA and CA Z-scan technique. When excited with 532 nm ns laser, we observed strong intensity-dependent ESA ($\beta_{ESA}$) while free-carrier absorption induced large positive $n_2$ in NCO. On the other hand, for 800 nm fs laser excitation, intensity-independent TPA is induced in NCO, while bound carriers gives rise to negative $-n_2$. We explain the observed nonlinearity by considering band structure of NCO and by taking into consideration the inter-band and intra-band transitions between VB and CB, consists of high and low spin states. Since both $\beta$ and $n_2$ obtained for NCO is comparable with well-known established TMO materials, we demonstrate its straightforward application in optical limiting devices.

## Acknowledgments

The authors gratefully acknowledge the Science and Engineering Research Board (project no. CRG/2019/002808), DAE BRNS (sanction no. 37(3)/14/26/2016-BRNS/37245), and FIST Project for Department of Physics.

**Table 1**: XPS survey spectra of NCO.

| | Species | | | | | |
|---|---|---|---|---|---|---|
| | **Ni 2p** | | **Co 2p** | | **O 1s** | |
| | $Ni^{2+}$ | $Ni^{3+}$ | $Co^{2+}$ | $Co^{3+}$ | Unsaturated O ion | -OH group |
| Binding Energy (eV) | 854.48 | 855.93 | 780.86 | 779.16 | 531.50 | 532.89 |
| | 871.56 | 872.80 | 796.41 | 794.38 | | |

**Table 2**: Comparison of nonlinear absorption coefficient (β) and nonlinear refractive index ($\eta_2$) with previously reported samples when excited with 532 nm nanosecond laser.

| Sample | Pulse duration | β (cm/GW) | $\eta_2$ (cm$^2$/GW) | Ref. |
|---|---|---|---|---|
| $NiCo_2O_4$ | 7 ns | 54.9 | 1.8 x 10$^{-4}$ | Present work |
| $ZnCo_2O_4$ | 7 ns | 27.4 | --- | 32 |
| α-$NiMoO_4$ | 7 ns | 71 | --- | 33 |
| ZnO | 5 ns | 7.6 | --- | 34 |
| $Fe_2O_3$ | 15 ns | 1.0 | --- | 35 |
| $Cr_2O_3$ | 4 ns | 3.17 | --- | 36 |

**Table 3**: Optical nonlinear parameters $I_{sat}$, $\beta$, and $n_2$ of NCO at 800 nm, 120 fs laser excitation.

| Sample | Intensity (GW/cm²) | $I_s$ (GW/cm²) x 10² | $\beta_{TPA}$ (cm/GW) x 10⁻² | $\eta_2$ x 10⁻⁷ (cm²/GW) |
|---|---|---|---|---|
| NCO | 47.3 | --- | --- | -14.32 ± 0.04 |
| | 63 | 4.0 ± 1.0 | 3.0 ± 0.6 | --- |
| | 78.9 | 3.0 ± 0.2 | 3.4 ± 0.3 | -0.780 ± 0.002 |
| | 94.6 | 2.7 ± 0.1 | 3.3 ± 0.1 | ---- |
| | 110.4 | 3.0 ± 0.2 | 3.0 ± 0.2 | -0.05 ± 0.001 |

**Table 4**: Comparison of nonlinear absorption coefficient (β) and nonlinear refractive index ($\eta_2$) with previously reported samples when excited with 800 nm femtosecond laser.

| Sample | Pulse Duration | β x $10^{-2}$ (cm/GW) | $\eta_2$ x $10^{-7}$ ($cm^2$/GW) | Ref. |
|---|---|---|---|---|
| $NiCo_2O_4$ | 120 fs | 3.4 | -14.32 | Present work |
| CuO | 60 fs | 1.0 | --- | 38 |
| $Fe_2O_3$ | 100 fs | 8.2 x $10^{-2}$ | --- | 39 |

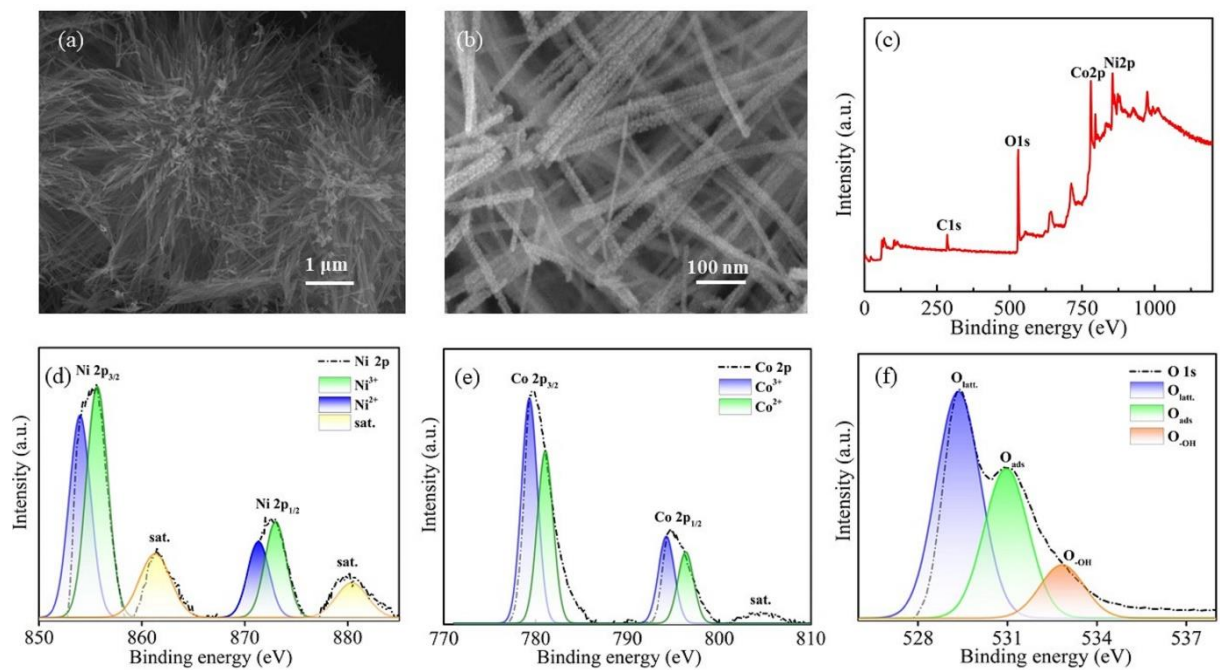

**Figure 1**. (a) Low and (b) high magnification FESEM images of NCO. The scale bars are given inside the images. XPS spectra of (c) survey spectrum, (d) Ni 2p, (e) Co 2p, and (f) O 1s for NCO.

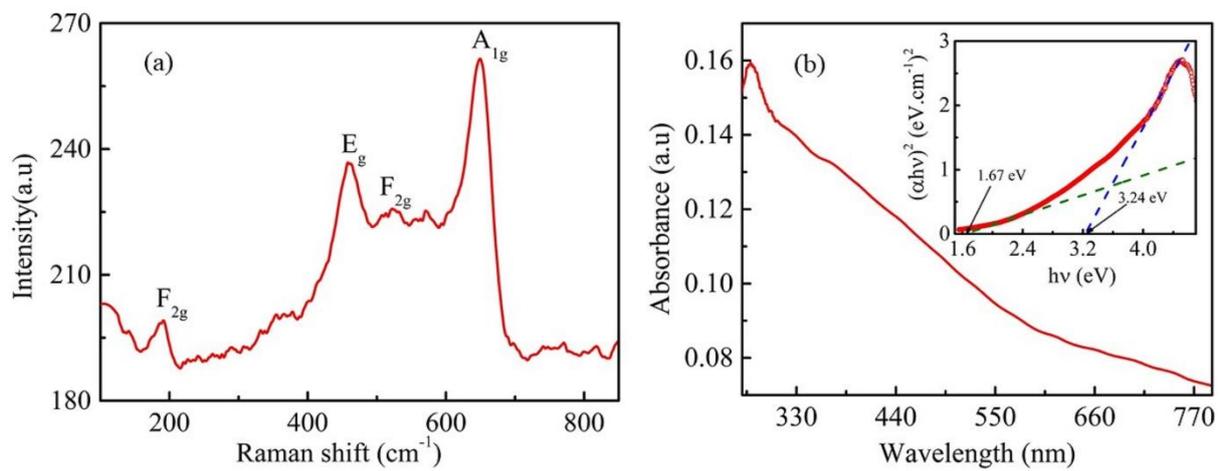

**Figure 2**. (a) Raman spectrum showing different vibrational modes in NCO. (b) Optical absorption spectra of NCO dispersed in NMP at 0.1 mg/mL . The inset represents the Tauc plot for calculating optical bandgap.

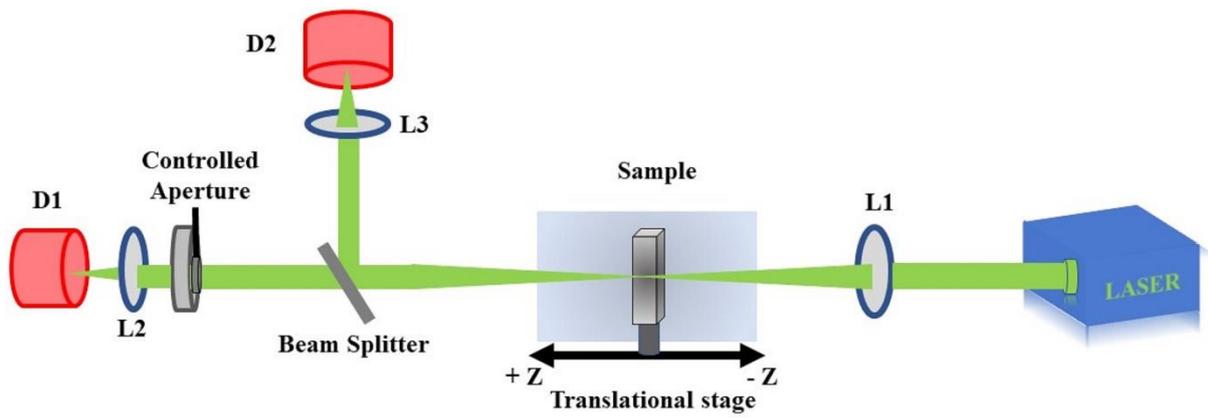

**Figure 3**. Schametic diagram of standard OA and CA Z-scan set-up. L and D represent lens and detectors, respectively.

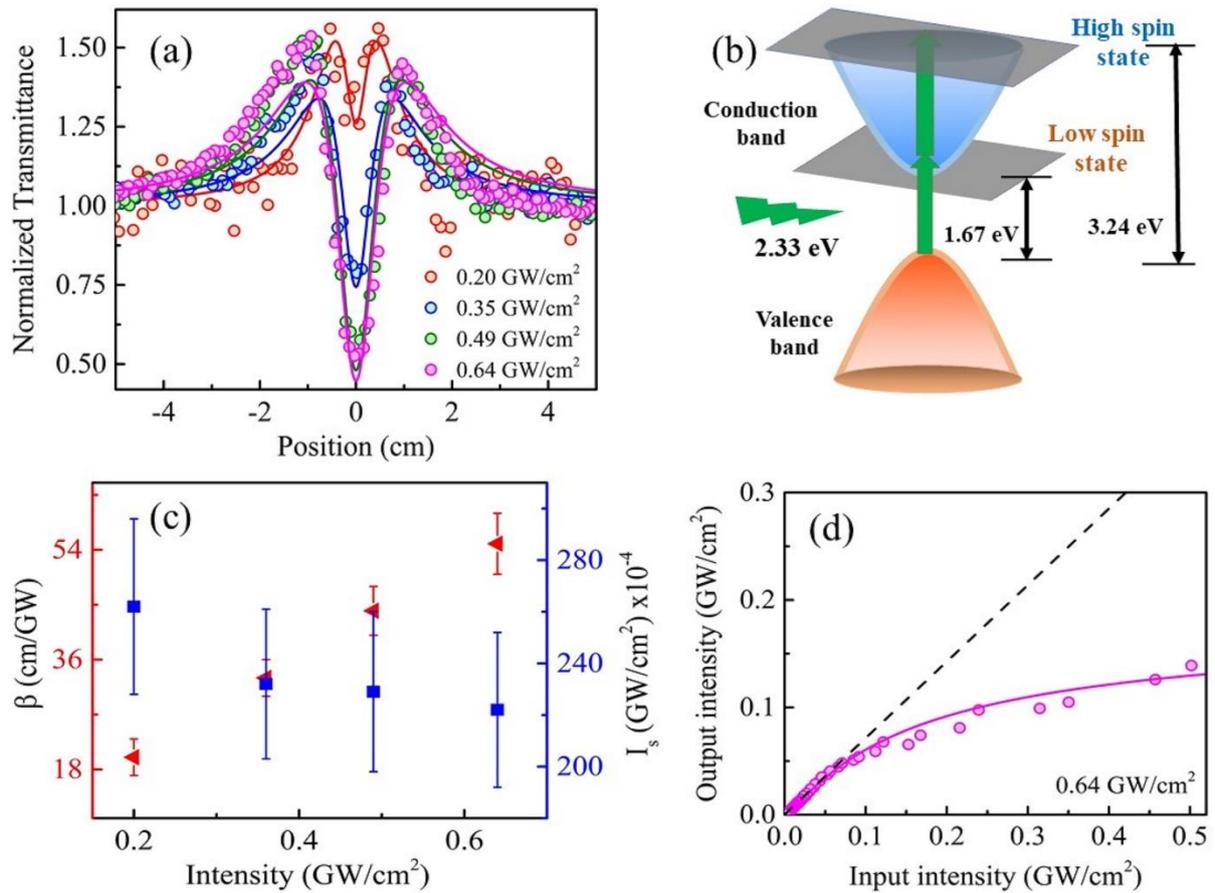

**Figure 4**. (a) Open aperture Z-scan traces of NCO at different pumping intensities when excited with 7 ns, 532 nm laser pulses. (b) Schametic diagram demonstrating electronic transition under ns illumination. (c) Variation of $β_{ESA}$ coefficient and $I_s$ as a function of input intensity. (d) Variation of output intensity as a function of input intensity for NCO at the peak intensity of 0.64 GW/cm$^2$.

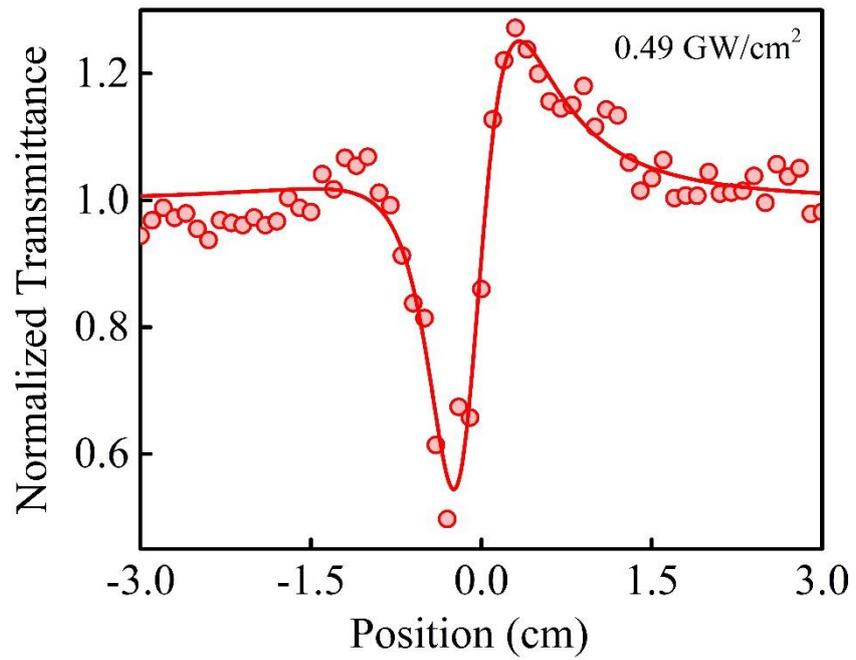

**Figure 5**. Closed-aperture Z-scan traces of NCO at 0.49 GW/cm$^2$. Hollow circles and solid line represent experimental data and theoretical fit, respectively.

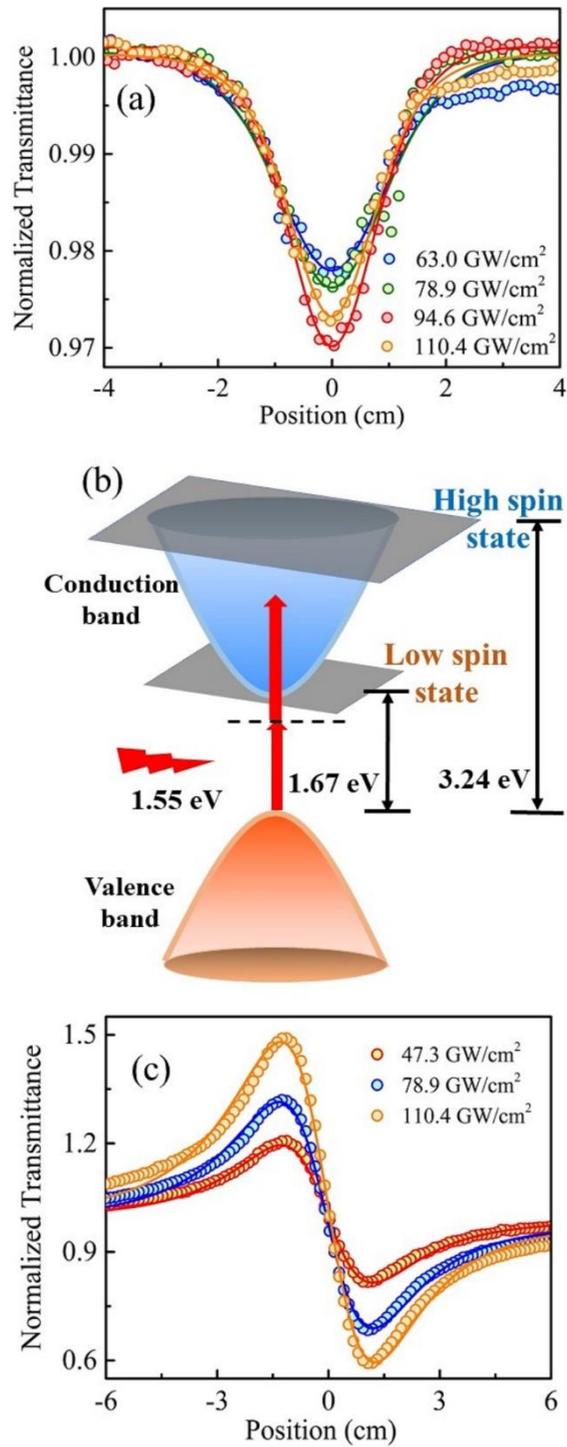

**Figure 6**. (a) Open aperture Z-scan traces of NCO at different pumping intensities when excited with 120 fs, 800 nm laser pulses (b) Schametic diagram demonstrating electronic transition under ns illumination. (c) Closed aperture Z-scan traces of NCO at different pumping intensities.